\documentclass[12pt]{iopart}

\usepackage{graphicx}
\usepackage{iopams}
\begin{document}

\title[]{Experimental proposal for measuring the Gouy phase of matter waves}

\author{I. G. da Paz$^{1,2}$, P. L. Saldanha$^{2,3}$, M. C. Nemes$^{2}$ and J. G. Peixoto de Faria$^{4}$}

\address{$^1$ Departamento de F\'{\i}sica, Universidade Federal do Piau\'{\i},
Campus Ministro Petr\^{o}nio Portela, CEP 64049-550, Teresina, PI,
Brazil}
\address{$^2$ Departamento de F\'{\i}sica,
Instituto de Ci\^{e}ncias Exatas, Universidade Federal de Minas
Gerais, CP 702, CEP 30161-970, Belo Horizonte, Minas Gerais, Brazil}
\address{$^3$ Department of Physics, University of Oxford,
Clarendon Laboratory, Oxford, OX1 3PU, United Kingdom}
\address{$^4$ Departamento de F\'{\i}sica e Matem\'{a}tica, Centro
Federal de Educa\c{c}\~{a}o Tecnol\'{o}gica de Minas Gerais, Av.
Amazonas 7675, CEP 30510-000, Belo Horizonte, MG, Brazil}

\ead{irismar@fisica.ufmg.br}

\begin{abstract}
The Schr\"odinger equation for an atomic beam predicts that it must
have a phase anomaly near the beam waist analogous to the Gouy phase
of an electromagnetic beam. We propose here a feasible experiment
which allows for the direct determination of this anomalous phase
using Ramsey interferometry with Rydberg atoms. Possible
experimental limitations are discussed and shown to be completely
under control within the present day technology. We also discuss how
this finding can open the possibility to use the spatial mode
wavefunctions of atoms as q-dits, since the Gouy phase is an
essential ingredient for making rotations in the quantum states.
\end{abstract}


\maketitle
\section{Introduction}
As early as 1890, Gouy \cite{gouy1,gouy2} has demonstrated the
existence of an anomalous phase, which carries his name, Gouy phase,
using classical electromagnetic fields in the paraxial wave regime.
This anomaly exists for any wave, including acoustic waves, and
results in a phase advance of $\pi$ for the amplitude of a Gaussian
beam after passing through the focus, propagating from $-\infty$ to
$+\infty$, in relation to a plane wave. This phase anomaly has
important consequences in optics and has been highly exploited in
recent experiments
\cite{Kandpal,Ruffin,feurer,Lindner,Thijs,zhu,Kawase}.

It was soon realized  that the paraxial wave equation for light is
formally equivalent to the 2-dimensional Schr\"odinger equation for
a free particle \cite{Berman,Monika,Paz1,Paz2}. This implies that
the Gouy phase, which is present in the solutions of the paraxial
equation, must also be present in the solutions of the Schr\"odinger
equation for a free particle. Recently we have shown \cite{Paz1}
that the Gouy phase for matter waves is compatible with diffraction
experiments made with fullerene molecules \cite{Zeilinger1} by
noting that the matter counterpart of the Gouy phase is directly
related to the covariance $\Delta
xp=\langle\hat{x}\hat{p}+\hat{p}\hat{x}\rangle/2$, where $\hat{x}$
and $\hat{p}$ are the position and momentum operators for the
molecules center of mass in the $x$ direction. However, this is an
indirect evidence, and a direct measurement of the Gouy phase for
matter waves might prove to be an interesting research tool.
And with the experimental advance that is actually in progress, the
Gouy phase may soon acquire the same importance for matter waves as
it already has for electromagnetic waves.

The paraxial Helmholtz equation describes the behaviour of the
amplitude $A(\mathbf{r})$ of an electromagnetic wave written as
$E(\mathbf{r},t)=A(\mathbf{r})\mathrm{exp}(\rmi kz-\rmi\omega t)$
when $A(\mathbf{r})$ varies slowly with $z$ such that $\partial^2
A/\partial z^2$ may be disregarded in relation to $k\partial
A/\partial z$. With this consideration, the wave equation gives
\cite{Saleh}
\begin{equation}
    \left[ \frac{\partial^2}{\partial x^2} +\frac{\partial^2}{\partial y^2} +2\rmi k\frac{\partial}{\partial z}\right]A(x,y,z)=0.
\end{equation}
The above equation is completely equivalent to the 2-dimensional
Schr\"odinger equation for a free particle if we make the
substitutions $k\rightarrow m/\hbar$ and $z\rightarrow t$, where $m$
is the mass of the particle, with the particle wavefunction
$\psi(x,y,t)$ in the place of $A(x,y,z)$. One particular solution of
the above equation is the Gaussian beam in the $x$ direction, where
we disregard the behaviour of the beam in the $y$ direction
\cite{Saleh}
\begin{equation}\label{gausbeam}
    A_G(x,z)=A_0 \frac{w_0}{w(z)}\mathrm{exp}\left[-\frac{x^2}{w(z)^2}+ \rmi\frac{kx^2}{2R(z)} - \rmi\xi(z)  \right],
\end{equation}
where the beam width $w(z)$, the curvature radius $R(z)$, the Gouy
phase $\xi(z)$ and the Rayleigh range $z_0$ are given by
\begin{eqnarray}\nonumber
&&w(z)=w_0\sqrt{1+\left(\frac{z}{z_0}\right)^2},\;R(z)=z\left[1+\left(\frac{z_0}{z}\right)^2\right],\\
&&\xi(z)=\frac{1}{2}\arctan\left(\frac{z}{z_0}\right)\;\mathrm{and}\;z_0=\frac{k
w_0^2}{2}\;.
\end{eqnarray}
The total variation of the Gouy phase when we go from $z=-\infty$ to
$z=+\infty$ with this beam is $\pi/2$. If the $y$ dependence of the
amplitude is the same as the $x$ dependence, the Gouy phase for the
$y$ confinement is summed to the phase due to the $x$ confinement
and a total phase of $\pi$ is obtained. However, in our proposal the
atomic beam, that will have a behaviour analogous to an
electromagnetic beam, will be confined only in the $x$ direction. We
will also consider that the energy associated with the momentum of
the atoms in the $z$ direction is very high, such that we can
consider a classical movement of the atoms in this direction, with
the time component given by $t=z/v_{z}$. So the $x$ component of the
atomic wave function will have the same form as \eref{gausbeam} with
the substitutions $k\rightarrow m/\hbar$ and $z\rightarrow z/v_{z}$.

\section{Experimental proposal}
Atoms interacting with electromagnetic fields can suffer mechanical
effects, like deflections or deviations of the motion of their
center of mass \cite{Askin,Gordon}. This property is useful in
atomic optics, since it allows for the construction of devices that
focus matter beams \cite{Paz2,Ashkin78,Schleich1,Schleich2}, in an
analogous way to the focalization of light beams by ordinary lenses.
In wave optics, it is well known that the fields suffer the Gouy
phase shift in the focus region, like the Gaussian beam described
above around $z=0$. In a previous working we explicitly showed that
the same anomaly should occur around the focus of an atomic beam
\cite{Paz2}. In order to experimentally observe this effect we
propose an experiment with a focused Gaussian atomic beam. We will
use a cylindrical focusing in the $x$ direction, without changing
the beam wavefunction in the $y$ direction, what makes the total
Gouy phase be $\pi/2$.
\begin{figure}[htp]
\centering
\includegraphics[width=9.0 cm]{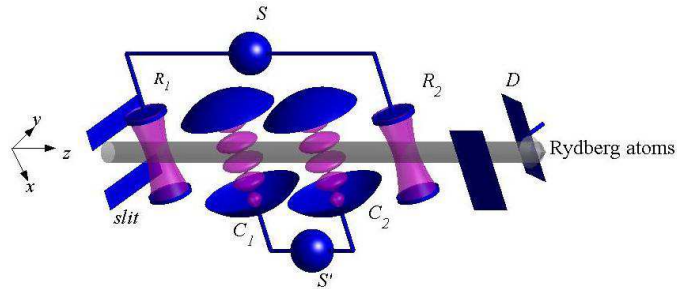}\\
\caption{Sketch of the experimental setup to measure the Gouy phase
for matter waves. Rydberg atoms are sent one-by-one with
well-defined velocity along the $z$-axis. A slit is used to
collimate the atomic beam in the $x$-direction. The Ramsey zones
$R_{1}$ and $R_{2}$ are two microwave cavities fed by a common
source $S$, whereas $C_{1}$ and $C_{2}$ are two high-Q microwave
cavities devised to work as thin lenses for the atomic beam. The
field inside these cavities is supplied by common source
$S^{\prime}$. The state of each atom is detected by the detector
$D$.}\label{figure1.eps}
\end{figure}

The experimental setup we propose to measure the Gouy phase shift of
matter waves is depicted in figure 1. This proposal is based on the
system of \cite{Haroche}. Rubidium atoms are excited by laser to a
circular Rydberg state with principal quantum number 49
\cite{Paulo,Gallagher}, that will be called state $|i\rangle$, and
their velocity on the $z$ direction is chosen to have a fixed value
$v_{z}$. As it was stated before, we will consider a classical
movement of the atoms in this direction, with the time component
given by $t=z/v_{z}$. A slit is used to prepare a beam with small
width in the $x$ direction, but still without a significant
divergence, such that the consideration that the atomic beam has a
plane-wave behaviour is a good approximation.
\begin{figure}[htp]
\centering
\includegraphics[width=4.0 cm]{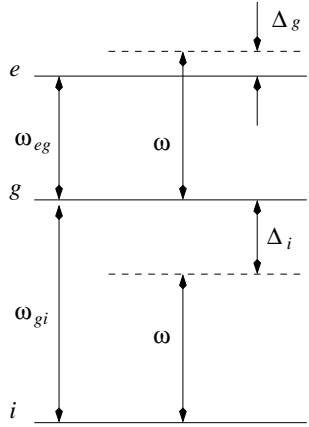}\\
\caption{Atomic energy levels compared with the frequency of the
field inside the cavities $C_{1}$ and $C_{2}$.}\label{figure2.eps}
\end{figure}
The relevant atomic states for the experiment, $|g\rangle$,
$|e\rangle$ and $|i\rangle$, along with the transition frequencies
among them, are illustrated in figure 2. If we disregard the
cavities $C_1$ and $C_2$, the setup is that of an atomic Ramsey
interferometer \cite{Ramsey}. The cavity $R_1$ has a field resonant
or quasi resonant with the transition
$|i\rangle\Leftrightarrow|g\rangle$ and results in a $\pi/2$ pulse
on the atoms, that exit the cavity in the state
$(|i\rangle+|g\rangle)/\sqrt{2}$ \cite{Haroche,Ramsey,Kim}. After
passing through the cavity $R_1$, the atoms propagate freely for a
time $t$ until the cavity $R_2$, that also makes a $\pi/2$ pulse on
the atoms. Calling $\hbar\omega_g$ and $\hbar\omega_i$ the energy of
the internal states $|g\rangle$ and $|i\rangle$ respectively,
$\omega_r$ the frequency of the field in the cavities $R_1$ and
$R_2$ and defining $\omega_{gi}\equiv \omega_g-\omega_i$, the
probability that detector $D$ measures each atom in the $|g\rangle$
state is
 $
P_0=\cos^2[(\omega_r-\omega_{gi})t]$ \cite{Haroche,Ramsey,Nogues}.
Upon slightly varying the frequency $\omega_r$ of the fields in
cavities $R_1$ and $R_2$, the interference fringes can be seen
\cite{Haroche,Ramsey,Nogues}.

The cavities $C_1$ and $C_2$ act as selective atomic lenses. Our
description of atomic lenses is based in \cite{Schleich1}. The
frequency of the fields in these cavities $\omega$ is supposed to be
strongly detuned from the atomic transition frequencies, not
changing the internal states of the atoms with their passage. In
this case we can describe the interaction by an effective
Hamiltonian \cite{Schleich1} $
H_{eff}^{(n)}=-\alpha_n|\mathbf{E}(\mathbf{r})|^2$
where $\mathbf{E}(\mathbf{r})$ represents the electric field in the
cavity and $\alpha_n$ is the atomic linear susceptibility, that
depends on the internal state $|n\rangle$ of the atom
\begin{equation}
    \alpha_n=\sum_m\frac{|\langle    m|\hat{\mathbf{P}}|n\rangle|^2}{\hbar}\left[\frac{1}{(\omega_m-\omega_n)+\omega}+\frac{1}{(\omega_m-\omega_n)-\omega}\right]\;,
\end{equation}
where $\hbar\omega_n$ is the energy of the level $|n\rangle$ and
$\hat{\mathbf{P}}$ an operator that corresponds to the atomic
electric dipole moment. When the field frequency $\omega$ is much
closer to one particular difference $\omega_m-\omega_n$ (but not
sufficiently close to induce a transition), we can consider only the
last term in the square brackets in the sum above. So we can write
the atomic linear susceptibilities for the states $|i\rangle$ and
$|g\rangle$ as 
\begin{equation}
    \alpha_i=\frac{|\langle g|\hat{\mathbf{P}}|i\rangle|^2}{\hbar\Delta_i}\;,\;
    \alpha_g=\frac{|\langle e|\hat{\mathbf{P}}|g\rangle|^2}{\hbar\Delta_g}\;,
\end{equation}
with $\Delta_g\equiv (\omega_e-\omega_g)- \omega$ and
$\Delta_i\equiv (\omega_g-\omega_i)- \omega$. If we choose a
frequency $\omega$ such that $|\Delta_g|\ll|\Delta_i|$, it is
possible that the component with internal state $|g\rangle$ of the
atomic wavefunction suffers the influence of the cavity field, while
the component with internal state $|i\rangle$ does not.

If the field in the cavities $C_1$ and $C_2$ has an electric field
node in the axis of the atomic beam and the width of the atomic beam
is much smaller than the wavelength of the cavity field, we can
consider the following Hamiltonian for the $|g\rangle$ part of the
atomic wavefunction in the cavity, expanded up to the second order
in $x$ \cite{Paz2,Schleich1,Schleich2}
\begin{eqnarray}
H_{eff}^{(g)}(x)&=&-\alpha_g |\mathbf{E}_0|^2\sin^2(2\pi
x/\lambda)\simeq-\alpha_g
|\mathbf{E}_0|^2\left(\frac{2\pi}{\lambda}\right)^2x^2\nonumber\\
&\simeq&-\hbar\frac{N\Omega^2}{\Delta_g}\left(\frac{2\pi}{\lambda}\right)^2x^2\;,
    \label{potencial}
\end{eqnarray}
where $\lambda$ is the wavelength of the cavity field,
$|\mathbf{E}_0|^2$ corresponds to an effective square of the
electric field on an antinode of the cavity, that is an average of
$|\mathbf{E}(\mathbf{r})|^2$ on the $z$ position inside the cavity
in an antinode, $\Omega=|\langle
g|\hat{\mathbf{P}}|i\rangle\cdot\mathbf{E}_0|/\hbar$ is the Rabi
frequency per photon of the system multiplied by $2\pi$  and $N$ is
the number of photons in each cavity. Let us disregard the kinetic
energy term of the Hamiltonian inside the cavity \cite{Schleich1}.
If the atoms enter the cavity with a wavefunction
$\psi_0(x)(|i\rangle+|g\rangle)/\sqrt{2}$ and interact with the
cavity field during an effective time $t_i$, the $|g\rangle$
component of the atomic wavefunction will exit the cavity as
\begin{equation}
    \psi'(x)|g\rangle=\rme^{-\rmi mx^2/(2\hbar t_{F})}\psi_0(x)|g\rangle\rme^{-\rmi\omega_g
    t_i},
\end{equation}
with
\begin{equation}
 \frac{1}{t_{F}}=-\frac{2\hbar t_i}{m}\frac{N\Omega^2}{\Delta_g}\left(\frac{2\pi}{\lambda}\right)^2.
\end{equation}

An optical converging cylindrical lens with focal distance $f$ puts
a quadratic phase $-kx^2/(2f)$ on the electromagnetic beam
\cite{Saleh}. By the analogy of the Schr\"odinger equation with the
paraxial Helmholtz equation \cite{Berman,Monika,Paz1}, we see that
the cavities act on the $|g\rangle$ component of the atomic beam as
cylindrical lenses with ``focal time'' $t_{F}$ and focal distance
$z_{F}=v_{z}t_{F}$. If we have $z_{F}=d/2$, where $d$ is the
distance between the cavities $C_1$ and $C_2$, the system will
behave like the illustration in figure 3. The cavity $C_1$ will
transform the $|g\rangle$ component of the wavefunction in a
converging beam with the waist on a distance $d/2$ (represented by
solid lines). After its waist, the beam will diverge until the
cavity $C_2$. The $|g\rangle$ component of the wavefunction on the
position of cavity $C_2$ will have the same width and the opposite
quadratic phase of the state $\psi'(x)$ above, so the cavity $C_2$
will transform the divergent beam in a plane-wave beam again. The
$|i\rangle$ component of the wavefunction, on the other hand,
propagates as a plane-wave beam all the time (represented by dashed
lines), as its interaction with the field of the cavities $C_1$ and
$C_2$ is considered to be very small.
\begin{figure}[htp]
\centering
\includegraphics[width=7.0 cm]{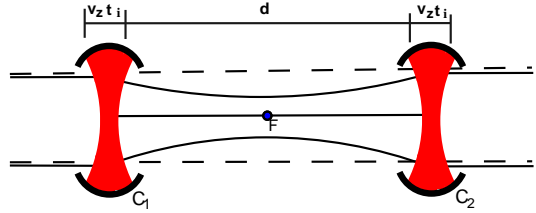}\\
\caption{Illustration of the operation of the cavities $C_{1}$ and
$C_{2}$ as thin lenses for the atomic beam. The dashed lines
represent the width of the atomic beam if the cavities are empty. If
a field is present, the solid lines represent the width of a beam
composed of atoms in the state $|g\rangle$. $F$ denotes the focus
region. On the other hand, if the beam is composed of atoms in the
state $|i\rangle$, the width does not change significatively.
}\label{figure3.eps}
\end{figure}

If we now repeat the Ramsey interference experiment, we will observe
a change of the positions of the fringes, because now the
$|g\rangle$ component acquires a $\pi/2$ Gouy phase due to the
cylindrical focusing that is not shared by the $|i\rangle$
component. So the interference pattern will be $
P'=\cos^2((\omega_r-\omega_{gi})t-\pi/2)\;.$
The difference on the positions of the minimums and maximums of the
patterns, one constructed when the field that forms the atomic
lenses is present on the cavities $C_1$ and $C_2$ and other when the
field is removed, should attest the existence of the Gouy phase for
matter waves.
\section{Experimental parameters and discussion}
As experimental parameters, we propose the velocity of the atoms
$v_{z}=50\;\mathrm{m/s}$ and a slit that generates an approximately
Gaussian wavefunction for the atoms
$\psi_0(x)\propto\rme^{-x^2/w_0^2}$ with $w_0=10\;\mathrm{\mu m}$.
The mass of Rubidium is $m=1.44\times10^{-25}\;\mathrm{kg}$. With
these parameters, the Rayleigh range of the atomic beam will be
$z_r=k_{z}w_{0}^{2}/2\simeq 3.5\;\mathrm{m}$ (where
$k_{z}=mv_{z}/\hbar$), much larger than the length of the
experimental apparatus, what justifies the plane-wave approximation.
On the cavities $C_1$ and $C_2$, we consider an interaction time
between the atoms and the atomic lenses $t_i=0.2\;\mathrm{ ms}$,
that corresponds to a width $v_{z}t_i=1\;\mathrm{cm}$ for the field
on the cavities. The wavelength of the field of the cavities $C_1$
and $C_2$ must be $\lambda\simeq5.8\;\mathrm{mm}$ \cite{Haroche},
with frequency near but strongly detuned from the resonance of the
transition $|g\rangle\Leftrightarrow|e\rangle$. The Rabi frequency
is about $\Omega/(2\pi)=47\;\mathrm{kHz}$ \cite{Haroche} and the
detuning chosen is $\Delta_g/(2\pi)=-30\;\mathrm{MHz}$, what makes
$\Delta_i/(2\pi)=+3.2\;\mathrm{GHz}$, such that with
$N=3\times10^{6}$ photons, an effectively classical field, the focal
distance for the atomic lenses is $10.5\;\mathrm{cm}$ for the
$|g\rangle$ component and $-11\;\mathrm{m}$ for the $|i\rangle$
component of the wavefunction. These parameters are consistent with
a separation of $d=21\;\mathrm{cm}$ between $C_1$ and $C_2$. All the
proposed parameters can be experimentally achieved
\cite{Haroche,Nogues,Gleyzes}.

The Rayleigh range $z_{r}^{\prime}$ and the beam waist
$w_{0}^{\prime}$ of the focused atomic beam can be calculated using
the analogy with the action of lenses in electromagnetic beams
considering that the incident beam has plane wavefronts
\cite{Saleh}: $z_{r}^{\prime}=z_r/[1+(z_{r}/z_F)^2]$,
$w'_0=w_0/\sqrt{1+(z_r/z_{F})^2},$
where $z_r$ and $w_0$ are the Rayleigh range and the beam waist of
the incident beam and $z_{F}$ is the focal distance of the atomic
lens. Using the proposed parameters, we have
$z_r'\simeq3\;\mathrm{mm}$ and $w'_0\simeq0.3\;\mathrm{\mu m}$. The
fact that $d\gg z_r'$ justifies our consideration that the
$|g\rangle$ component of the beam acquires a $\pi/2$ Gouy phase.

The interaction between the atomic beam and the field in the
cavities $C_1$ and $C_2$ depends on the position $x$, according to
\eref{potencial}. If we do not want that photons be absorbed by the
atoms, it is important that $4\pi^2 N\Omega^2 x^2/(\Delta_g^{2}
\lambda^2)\ll1$ for the entire beam \cite{Scully}. We have $4\pi^2
N\Omega^2w_0^2/(\Delta_g^{2}\lambda^2)\simeq8\times10^{-4}$ for the
proposed parameters, where $w_0$ is the beam width, showing that the
absorption of photons can be disregarded. It is also important that
the cavities $C_1$ and $C_2$ have a large quality factor $Q$. This
occurs because in \eref{potencial} it was considered that the
intensity of the electric field is exactly 0 in $x=0$, what is
impossible for real cavities. In fact, the ratio between the maximum
and the minimum of intensity in a cavity should roughly be the
quality factor $Q$. So the $|g\rangle$ component of the beam also
acquires a phase $N\Omega^2t_i/(\Delta_gQ)$ on the passage in each
cavity, and this phase will be added to the accumulated Gouy phase.
If we want that this undesired phase be smaller than $\pi/20$, we
need $Q>4\times10^{6}$ for our proposed parameters. This can also be
experimentally achieved \cite{Haroche,Nogues,Gleyzes}.

In our treatment for the atomic lenses, we disregarded the kinetic
energy term of the atoms in the Hamiltonian. To verify that this
assumption is in fact reasonable, we made the exact calculations for
the problem, considering the kinetic energy term in the Hamiltonian
following \cite{Schleich2}, section 20.4. The basic difference on
the results for the proposed parameters was a difference inferior to
2\% for the focal distance of the atomic lenses. So the disregarding
of the kinetic energy term does not represent a problem.

In our proposal, we used one particular combination of parameters
for the experiment within today's experimental capabilities. It is
important to stress that many of the proposed parameters can be
varied in a wide range, making it possible to choose the most
appropriate ones in an experiment.

The Gouy phase for matter waves could have important applications in
the field of quantum information. The transversal wavefunction of an
atom in a beam state can be treated not only as a continuous
variable system, but also as an infinite-dimensional discrete
system. The atomic wavefunction can be decomposed in
Hermite-Gaussian or Laguerre-Gaussian modes in the same way as an
optical beam \cite{Saleh}, which form an infinite discrete basis.
This basis was used, for instance, to demonstrate entanglement in a
two-photon system \cite{mair01}. However, it is essential for
realizing quantum information tasks that we have the ability to
transform the states from one mode to another, making rotations in
the quantum state. This can be done using the Gouy phase,
constructing mode converters in the same way as for light beams
\cite{allen92,beijersbergen93}. The mode converters can transform
any mode (Hermite-Gaussian or Laguerre-Gaussian) into another mode
of the same order. A mode converter is composed simply by two
cylindrical lenses, that can focus and collimate the beam in one
direction. As different modes in general acquire different Gouy
phases  through the focalization in one direction \cite{Saleh}, for
a beam in a superposition of modes phase differences between the
components can be included, and a combination of these converters is
sufficient for making transformations between any two modes of the
same order \cite{beijersbergen93}. So the Gouy phase may make it
possible to use atomic beams of a determinate order in quantum
information schemes as q-dits. Recently electron beams in
Laguerre-Gaussian modes had been constructed
\cite{uchida10,mcmorran11}, and the same technique of
\cite{mcmorran11}, which uses diffraction gratings to generate the
beams, could be used to generate atomic Laguerre-Gaussian beams to
be used in such schemes. It should be also possible to implement a
similar scheme with trapped atoms, since the Hermite-Gaussian modes
are the eigenstates of harmonic oscillators. In this case the
focalization of the wavefunction could be made turning
electromagnetic fields on and off.

\section{Conclusion}
We had proposed a feasible experiment to directly measure the Gouy
phase for matter waves using atomic Ramsey interferometry. The
experimental parameters necessary for the implementation were shown
to be accessible under the current technology. The verification of
the Gouy phase in matter waves has the possibility to generate a
great amount of development in atomic optics, in the same way as the
electromagnetic counterpart Gouy phase had contributed to
electromagnetic optics. For instance, it can be used to construct
mode converters for atomic beams and trapped atoms, with potential
applications in quantum information.

\ack We would like to thank Prof. P. Nussenzveig and Prof. J. R. R.
Leite by fruitful discussions on microwave cavities. This work was
in part supported by the Brazilian agencies CNPq, Capes and Fapemig.








\section*{References}

\begin{thebibliography}{10}

 \bibitem{gouy1} Gouy L G 1890  {\it C. R. Acad. Sci. Paris} \textbf{110} 1251.

\bibitem{gouy2} Gouy L G 1891 {\it Ann. Chim. Phys. Ser. 6} \textbf{24}
145.

\bibitem{Kandpal} Kandpal H C, Raman S and Methrotra R 2007 {\it Optics
and Lasers in Eng.} \textbf{45} 249.

\bibitem{Ruffin} Ruffin A B, Rudd J V, Whitaker J F, Feng S and
Winful H G 1999 {\it Phys. Rev. Lett.} \textbf{83} 341.

\bibitem{feurer} Feurer T, Stoyanov N S, Ward D W and Nelson K A 2002
{\it Phys. Rev. Lett.} \textbf{88} 257402.

\bibitem{Lindner} Lindner F \etal 2004 {\it Phys. Rev. Lett.}
\textbf{92} 113001.

\bibitem{Thijs} Klaassen T, Hoogeboom A, van Exter M P and Woerdman J
P 2004 {\it J. Opt. Soc. Am.} A \textbf{21} 1689.

\bibitem{zhu} Zhu W, Agrawal A and Nahata A 2007 {\it Opt. Express}
\textbf{15} 9995.


\bibitem{Kawase} Kawase D, Miyamoto Y, Takeda M, Sasaki K and
Takeuchi S 2008 {\it Phys. Rev. Lett.} \textbf{101} 050501.





\bibitem{Berman} Berman P R 1997 {\it Atom Interferometry} (Academic
Press, San Diego).

\bibitem{Monika} Marte M A M  and Stenholm S 1997 {\it Phys. Rev.} A
\textbf{56}, 2940.

\bibitem{Paz1} da Paz I G, Nemes M C, Pádua S, Monken C H and Peixoto
de Faria J G 2010 {\it Phys. Lett.} A \textbf{374} 1660.

\bibitem{Paz2} da Paz I G, Nemes M C and  Peixoto de Faria J G 2007
{\it J. Phys.: Conference Series} \textbf{84} 012016.

\bibitem{Zeilinger1} Nairz O, Arndt M and Zeilinger A 2002 {\it Phys.
Rev.} A \textbf{65} 032109.


\bibitem{Saleh} Saleh B E A and Teich M C 1991 {\it Fundamentals of
 Photonics} (Jonh Wiley Sons, Inc., New York).


\bibitem{Askin} Ashkin A 1970 {\it Phys. Rev. Lett.} \textbf{24} 156.

\bibitem{Gordon} Gordon J P and Ashkin A 1980 {\it Phys. Rev.} A
\textbf{21} 1606.

\bibitem{Ashkin78} Bjorkholm J E, Freeman R R, Ashkin A and Pearson D
B 1978 {\it Phys. Rev. Lett.} \textbf{41} 1361.


\bibitem{Schleich1} Averbukh I S, Akulin V M and Schleich W P 1994
{\it Phys. Rev. Lett.} \textbf{72} 437.



\bibitem{Schleich2} Schleich W P 2001 {\it Quantum Optics in Phase
Space} (Wiley-VCH, Berlin).


\bibitem{Haroche} Raimond J M, Brune M and Haroche S 2001 {\it Rev. Mod.
Phys.} \textbf{73} 565.


\bibitem{Paulo} Nussenzveig P \etal 1993 {\it Phys. Rev.} A
\textbf{48} 3991.

\bibitem{Gallagher} Gallagher T F 1994 {\it Rydberg Atoms}
(Cambridge University Press, Cambridge).

\bibitem{Ramsey} Ramsey N F 1985 {\it Molecular Beams} (Oxford
University Press, New York).


\bibitem{Kim} Kim  J I \etal 1999 {\it Phys. Rev. Lett.} \textbf{82}
4737.





\bibitem{Nogues} Nogues G \etal 1999 {\it Nature} \textbf{400} 239.


\bibitem{Gleyzes} Gleyzes S \etal 2007 {\it Nature} \textbf{446} 297.

\bibitem{Scully} Scully M O and Zubary M S 1997 {\it Quantum Optics}
(Cambridge University Press, Cambridge).

\bibitem{mair01} Mair A, Vaziri A, Weihs G and Zeilinger A 2001
{\it Nature} \textbf{412} 313.

\bibitem{allen92} Allen L, Beijersbergen M W, Spreeuw R J C and
Woerdman J P 1992 {\it Phys. Rev.} A \textbf{45} 8185.

\bibitem{beijersbergen93} Beijersbergen M W, Allen L, van der Veen H
E L O and Woerdman J P 1993 {\it Opt. Comm.} \textbf{96} 123.


\bibitem{uchida10} Uchida M and Tonomura A 2010 {\it Nature} \textbf{464}
737.


\bibitem{mcmorran11} McMorran B J \etal 2011 {\it Science}
\textbf{331} 192.

\end{thebibliography}
\end{document}